\documentclass[manuscript]{aastex631}
\usepackage[utf8]{inputenc}
\usepackage{amsmath}
\usepackage{lipsum}
\usepackage{color,soul}
\usepackage{appendix}
\usepackage{tikz}
\usepackage{hyperref}

\begin{document}

\title{Exploring the Connection between Helioseismic Travel Time Anomalies and the Emergence of Large Active Regions during Solar Cycle 24}

\correspondingauthor{John T. Stefan}
\email{jts25@njit.edu}

\author[0000-0002-5519-8291]{John T. Stefan}
\affiliation{Department of Physics, New Jersey Institute of Technology, Newark, NJ 07102}

\author[0000-0003-0364-4883]{Alexander G. Kosovichev}
\affiliation{Department of Physics, New Jersey Institute of Technology, Newark, NJ 07102}
\affiliation{NASA Ames Research Center, Moffett Field, Mountain View, CA 94040}

\begin{abstract}
    We investigate deviations in the mean phase travel time of acoustic waves preceding the emergence of 46 large active regions observed by the Helioseismic and Magnetic Imager (HMI). In our investigation, we consider two different procedures for obtaining the mean phase travel time, by minimizing the difference between cross-correlations and a reference, as well as the Gabor wavelet fitting procedure. We cross-correlate the time series of mean phase travel time deviations with the surface magnetic field and determine the peak correlation time lag. We also compute the perturbation index--the area integrated mean phase travel time deviations exceeding quiet sun thresholds--and compare the time of peak perturbation index with the correlation time lag. We find that the lag times derived from the difference minimization procedure precede the flux emergence for 36 of the 46 active regions, and that this lag time has a noticeable correlation with the maximum flux rate. However, only 28 of the active regions have peak perturbation index times in the range of 24 to 48 hours prior to the flux emergence. Additionally, we examine the relationship between properties of the emerged active regions and the strength of helioseismic signals prior to their emergence.
\end{abstract}

\section{Introduction}

With the increase of interest in space weather forecasting, there is a strong need for better understanding of the development of solar active regions and the origin of their magnetic flux. There have been several notable attempts to detect magnetic flux as it emerges from the solar interior, falling into two types of analyses. The first type examines changes in the surface acoustic power prior to the emergence of magnetic flux, as in \citet{Hartlep} where simulations of submerged sound speed perturbations are analyzed. The authors find a frequency-dependent reduction in the acoustic power, with a maximum reduction of $35\%$ in the 2-4 mHz range as compared to the unperturbed areas. Conversely, \citet{Singh1} find enhancement of the acoustic power in the same range using different methodology. The authors use ring-diagrams to isolate the $f$-mode from the power spectrum and find that for several active regions, the power is reduced one to two days prior to the flux emergence. Recent study of the $f$-mode technique, however, has yielded inconclusive results. \citet{Waidele} examine the $f$-mode using a Fourier-Hankel decomposition of the surface wave field instead of the typical ring diagram, and the strengthening of the $f$-mode prior to active region emergence is again observed. On the other hand, \citet{Korpi} analyze the same active regions with additional calibration steps, and find no significant enhancement of the $f$-mode.

Aside from examining changes in the surface acoustic power, other analyses have looked for direct evidence of the emerging magnetic flux. In particular, \citet{Stathis2013} identified significant reductions in the travel time of oscillations on the order of 10 seconds one to two days prior to the emergence of several active regions. Similar work by the authors in a previous publication \citep{Stathis_nature} was contested by \citet{Braun_comment}, who was unable to reproduce the reported travel time deviations using helioseismic holography. In their reply, \citet{Ilonidis_response} state that the discrepancy likely comes from differences in filtering and measurement techniques. However, the results of \citet{Stathis_nature} were confirmed by \citet{Kholikov} who found similar 10 to 15 second reductions in the mean phase travel time at depths of 40 to 75 Mm. Additional investigation at shallower depths, between 2 and 15 Mm beneath the photosphere, by \citet{Toriumi} of an active region studied by \citet{Stathis2013} also confirm mean phase travel time reductions preceding the active region's emergence. Further study of the same active region also identified phase travel time reductions preceding emergence \citep{StefanAR}, and a comprehensive analysis of 100 active regions using helioseismic holography also found similar signatures \citep{Birch_stats}.

There is a clear need for additional examination of these pre-emergence signatures. One of the main critiques of mean travel time analyses is the large difference between the 10 to 15 second variations which were measured and the theoretical effect of the magnetic field in the convection zone, which should produce immeasurably small variations in the travel time \citep{Braun_comment}. A recent study by \citet{Felipe} may provide an alternative explanation, where the authors perform MHD simulations to develop a realistic sunspot. The authors then initiate two simulations, one with the previous sunspot's thermal profile with zero magnetic field, and another with the previous sunspot's magnetic field but with a quiet sun thermal profile, with the goal of identifying the individual contribution of thermal and magnetic effects. While the magnetic-only sunspot produces weak mean travel time shifts, those from the thermal simulation are both stronger and more persistent with depth. Furthermore, the thermal variations should decrease the wave speed and lengthen the travel time, but strong reductions in the travel time are measured instead. The authors show that this can be explained by considering the corresponding change in the cut-off frequency, which would shorten the travel time in agreement with \cite{Moradi}. It may be possible that similar thermal effects originating from the concentration of rising magnetic flux can lead to the observed pre-emergence travel time variations.

Our goal here is to determine the significance of these variations and examine their relationship with other active region characteristics. Here, we follow a similar helioseismic procedure as outlined by \cite{Stathis2013} and \citet{StefanAR} for the analysis of several active regions observed by the Helioseismic and Magnetic Imager (HMI) during Solar Cycle 24. We restrict the analysis only to sufficiently large active regions, where any possible pre-emergence signatures should be measurable.

\section{Methods}

We begin first by searching the NOAA Solar Region Summaries catalog for active regions with maximum size greater than or equal to 150 millionths of disk area, appearing between April 2011 and November 2020. The resulting list is further refined, selecting only active regions which are first identified within 45 heliographic degrees from disk center. We then obtain the tracked Dopplergram and Magnetogram series for each of these active regions, 91 in total, as the initial dataset, with each time series beginning 72 hours prior to the identification time and ending 24 hours after the identification time. The Dopplergrams and Magnetograms both have spatial resolution of 0.12 degrees per pixel, with the Dopplergrams having cadence 45 seconds and the Magnetograms having cadence 12 minutes, and the total spatial extent of both series are 30.72 by 30.72 heliographic degrees. Next, the Dopplergrams are examined for any artifacts which may interfere with their processing, such as missing frames in the series or anomalous pixels, and 20 of the initial 91 active region Dopplergram series were found to be unsuitable for analysis. The remaining 71 Dopplergram series are subdivided into 8-hour segments in increments of 4 hours for our analysis, and we present the results of the first 46 in this work.

Each Dopplergram segment is treated with a Fourier filter, as in \citet{StefanAR}, to isolate oscillations which have a turning point between 40 Mm and 70 Mm beneath the photosphere. First, oscillations with frequencies outside of the resonant portion of the spectrum above 5 mHz, as well as those lying in the range of convective noise below 2 mHz, are removed. A flat-top phase speed filter is then applied, where oscillations above $v_p=\omega/k_h=92$ km/s and below $v_p=127$ km/s are left untreated, and outside of this range treated with a Gaussian filter of width $8.7$ km/s centered at the corresponding phase speed.

\begin{figure}
\begin{center}
    \includegraphics[width=0.75\linewidth]{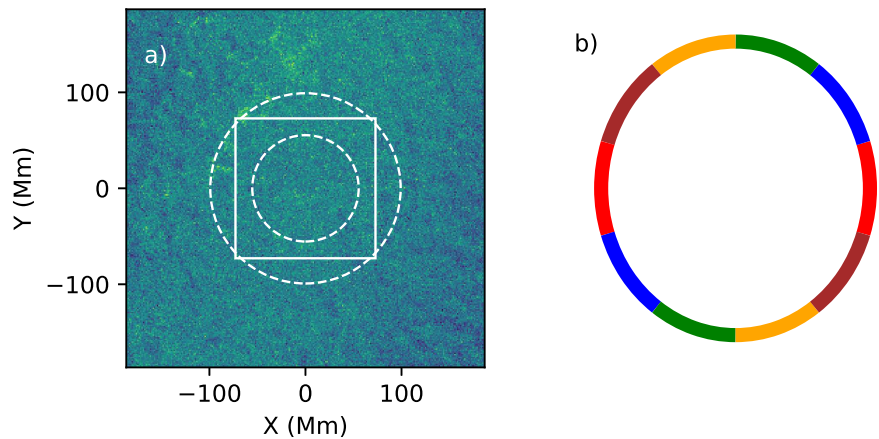}
    \caption{a) Acoustic power map for a quiet sun region, with the resulting map size (solid square) and smallest and largest measurement annuli (dashed circles) overlayed. b) Example of annulus division for the case of 10 arc segments.}\label{example}
\end{center}
\end{figure}

We next compute the cross-correlation of the filtered Dopplergram signal for a variety of measurement schemes. The basis of each scheme is the same, where the signal is averaged within sectors of 31 annuli of uniformly increasing radius, with minimum radius 4.56 degrees and maximum radius 8.16 degrees corresponding to the turning points at the depths of 40 Mm and 70 Mm, respectively. The size of these annuli, as well as the size of our output map, are displayed in Figure \ref{example}a. We consider five separate cases of the number of sectors, ranging from 6 to 14 sectors in increments of 2, and the Doppler signal within each sector is averaged and cross-correlated with the opposing sector; Figure \ref{example}b shows how an annulus is divided for the case of 10 sectors. The resulting cross-correlations are then used to determine the mean phase travel time using two distinct methods.

The first method fits the cross-correlations for the negative and positive time lags to a Gabor wavelet, as described in \citet{Gabor_nigam}, with the amplitude, frequency, phase travel time, group travel time, and width of wave packet as the fitting parameters. The positive and negative phase travel times are averaged to obtain the mean phase travel time, and the quiet sun mean phase travel time is subtracted to obtain the deviation. The second method, developed by \citet{GB02} and referred to as the GB02 method, determines the mean phase travel time deviation directly by comparing a given cross-correlation with some reference. Here, we choose quiet sun cross-correlations obtained from performing the above procedure on Dopplergrams of regions with no significant magnetic features. The positive and negative phase travel time deviations are identified as the time lags which minimize the squared difference between a given cross-correlation and the reference. Again, the two deviations are averaged to obtain the mean phase travel time deviation.

With the resulting travel time deviation maps, we consider two quantities in evaluating the connection with the surface magnetic flux: the perturbation index and the correlation lag. The perturbation index is obtained by integrating, over area, the travel time deviations exceeding a given threshold to give a measure of perturbation strength over time \citep{Stathis2013}. Because the perturbation index relies only travel time deviation maps of a given moment, it has potential for forecasting applications; generally, we expect strong perturbation indices to be indicative of rising magnetic flux. We take into account the variation of travel time deviations depending on the number of arc segments used to compute the cross-correlations by identifying a threshold for each case. The 99$^{th}$ percentile of the travel time deviations of three quiet sun patches, of the same size as the active regions as well as the same duration, are used as the threshold for the respective arc segment cases. These thresholds range from $-7.1$ to $-10.2$ seconds for the GB02 travel times, and from $-7.1$ to $-8.5$ for the Gabor-wavelet derived travel times. Here, we define the flux emergence time as the peak in the second time derivative of the surface magnetic flux or, in other words, when the flux rate is increasing most rapidly.

The correlation lag time, on the other hand, requires the entire sequence of travel time maps and Magnetograms. For each pixel, we subtract the mean and normalize the mean phase travel time deviations and the unsigned surface magnetic field, and then compute the cross-correlation. The normalization ensures that an absolute correlation can be determined, i.e.\ a value between $-1$ and $1$. The cross-correlations between the travel times and magnetic field are then averaged for pixels where the unsigned surface magnetic field eventually reaches $100$ G or more, and we identify the correlation lag time from the maximum in the averaged cross-correlation.

\subsection{Toy Model for Flux Emergence}

\begin{figure}
\begin{center}
    \includegraphics[width=\linewidth]{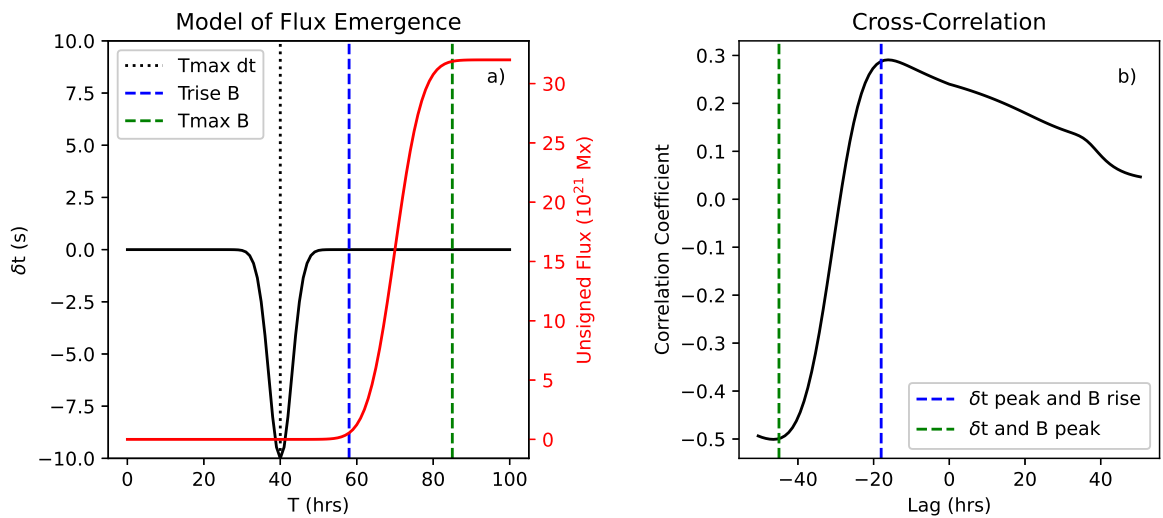}
    \caption{a) Simple model of flux emergence, with the magnetic field (red) increasing as an error function and the travel time deviation (black) peaking prior to emergence. b) Cross-correlation of the magnetic field and travel time deviation. The green dashed line in b) marks the minimum of the cross-correlation, with the lag equal to the separation of the travel time deviation and magnetic field peaks. The blue dashed line in b) marks the maximum of the cross-correlation, with the lag equal to the separation between the travel time deviation peak and onset of the magnetic field's rise.}\label{toy}
\end{center}
\end{figure}

In order to determine what features should be expected in the travel time - magnetic field cross-correlation, we consider a simple model for flux emergence. Here, the travel time deviation is Gaussian in time, with a FWHM of 8 hours and separated from the time of peak emergence by 30 hours. The magnetic field increases as an adjusted error function, taking around 20 hours to reach its peak. Both quantities are shown in Figure \ref{toy}a, with the vertical green dashed line indicating when the magnetic field has saturated and the blue dashed line showing the onset of flux emergence. The peak in the travel time deviation is indicated by the black dotted line, which serves as a reference point for comparison with the rise and peak times of the magnetic field.

We find that the minimum of the cross-correlation, indicated by the green dashed line in Figure \ref{toy}b, occurs at a lag equal to the time between the travel time deviation's peak and the increase of the magnetic field, i.e. the onset of flux emergence. Intuitively, we should expect the cross-correlation to peak here, as the return of the travel time deviation to zero has a similar functional form to the rise of the magnetic field. The maximum of the cross-correlation, indicated by the blue dashed line in Figure \ref{toy}b, occurs at a lag equal to the the time between the peak travel time deviation and peak magnetic field. This should also be expected, as the negative peak of the travel time deviation and the positive peak of the magnetic field contribute most strongly to the cross-correlation when the two are aligned, forming the indicated minimum. In this work, we are most interested in the delay between strong negative travel time deviations and the onset of flux emergence, so we will focus on the time lag associated with the minimum in the cross-correlation.

\section{Results}

\begin{figure}
\begin{center}
    \includegraphics[width=\linewidth]{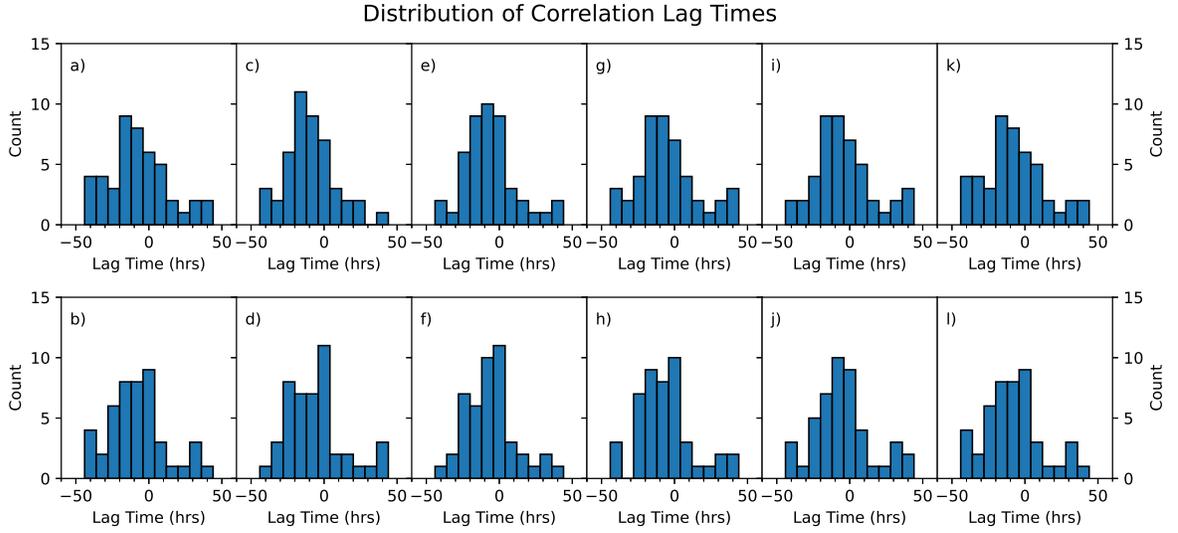}
    \caption{Distribution of correlation lag times for the travel times obtained with 6 arc segments (a,b), 8 arc segments (c,d), 10 arc segments (e,f), 12 arc segments (g,h), 14 arc segments (i,j), and by the average of the five cases (k,l). The first row is derived from the GB02 travel time procedure, and the second row from the Gabor wavelet fitting procedure.}\label{dist}
\end{center}
\end{figure}

We first examine the distribution of correlation lag times for each of the arc segment cases and travel time procedures, as shown in Figure \ref{dist}. In all cases, there are more negative lag times than positive, with the GB02-derived, 8 arc segment method yielding negative lag times in 36 of the 46 active regions; in fact, the GB02 procedure outperforms the Gabor wavelet procedure in every case. The median correlation lag time varies with the number of arc segments, though the 14 arc segment case performs most poorly with a median correlation lag time of $-10$ hours for the GB02 procedure and $-12$ hours for the Gabor wavelet procedure. The 8 arc segment case produces the greatest median correlation time lag of $-16$ hours for both procedures, and median correlation lag time decreases in both the GB02 and Gabor wavelet procedures as the number of arc segments increases.

\begin{figure}
    \centering
    \includegraphics[width=\linewidth]{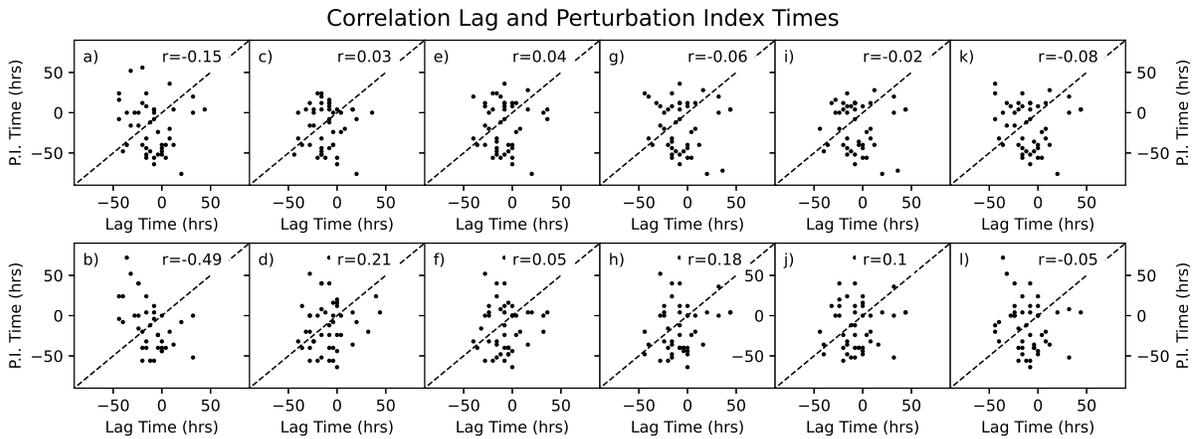}
    \caption{Comparison between the correlation lag times and the peak perturbation index times; the plots are organized as in Figure \ref{dist}.}
    \label{comp}
\end{figure}

The median lag time associated with the peak in the perturbation index does not follow this trend, however, and actually increases with the number of arc segments. The median lag time from the perturbation index in the 6 arc segment case is $-10$ hours for both travel time procedures, reaching a maximum of $-14$ hours for the GB02 procedure and $-20$ hours for the Gabor wavelet procedure, both with the 14 arc segment case. We next compare the perturbation index indicated lag time with the correlation lag time in Figure \ref{comp} and compute the Pearson correlation coefficient for each case. For the GB02 procedure, the two lag times are most correlated in the 8 arc segment case (Figure \ref{comp}e) with a weak correlation coefficient of $r=0.03$. The perturbation index and lag times derived from the Gabor wavelet procedure are also most correlated in the 8 arc segment case (Figure \ref{comp}d), with $r=0.21$. For both the GB02 and Gabor wavelet procedures, the greatest correlation between perturbation index and lag times are associated with the arc segment case which yields the greatest number of negative correlation lag times.

\begin{figure}
    \centering
    \includegraphics[width=0.5\linewidth]{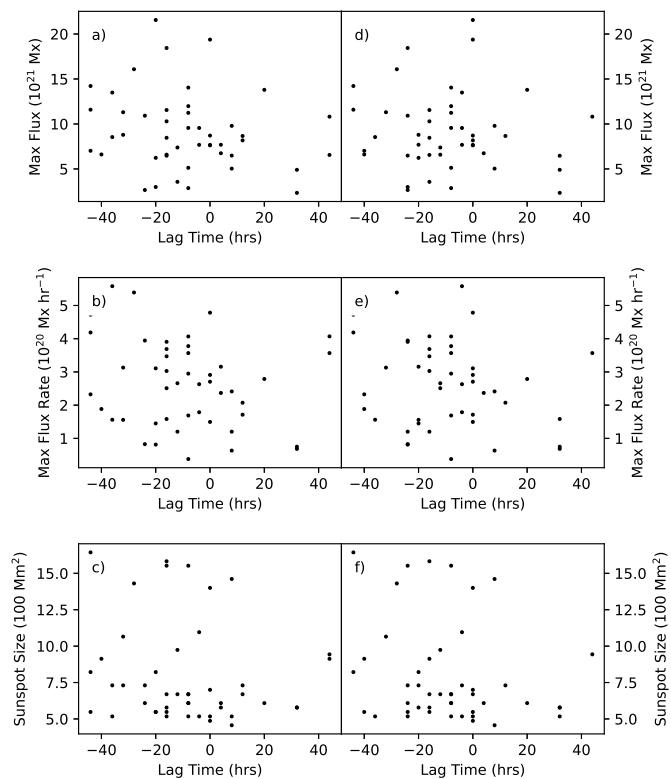}
    \caption{Comparison between the correlation lag times for the GB02 procedure (left column) and Gabor wavelet procedure (right column), and an active region's maximum magnetic flux (a,d); an active regions maximum flux rate (b,e); and an active region's size (c,f).}\label{chars}
\end{figure}

We now examine the relationship between the correlation lag times and relevant active region characteristics, and here we consider the case where the travel times are obtained through averaging the cross-correlations over the different number of arc segments. As opposed to the previous comparison, we do not necessarily expect a linear relationship with the active region characteristics, so here we use the Spearman correlation coefficient instead of the Pearson correlation coefficient. The Spearman coefficient evaluates the monotonicity of a relationship, where the Pearson coefficient evaluates the linearity. The first quantity we examine is the maximum magnetic flux of an active region, Figures \ref{chars}a and \ref{chars}d, and find that there is a weak anti-correlation with the correlation lag time for the GB02 procedure, and far weaker for the Gabor wavelet procedure, with $\rho=-0.22$ and $\rho=-0.09$ respectively. There is a relationship of similar strength between the correlation lag times and the maximum flux rates, Figures \ref{chars}b and \ref{chars}e, with $\rho=-0.40$ for both the GB02 and the Gabor wavelet procedures. There is also a weak anti-correlation between the correlation lag times and sunspot sizes, Figures \ref{chars}c and \ref{chars}f, with $\rho=-0.16$ and $\rho=-0.22$ for the GB02 and Gabor wavelet procedures, respectively.

After removing the active regions with positive correlation lag times, which do not predict flux emergence, the probability that the sunspot sizes are uncorrelated with the lag times increases, from $p=0.26$ to $p=0.39$ for the GB02 procedure and $p=0.14$ to $p=0.50$ for the Gabor procedure. Reevaluating the relationship between the maximum fluxes also leads to a significant increase in the probability of no correlation, increasing from $p=0.14$ to $p=0.40$ for the GB02 procedure and from $p=0.57$ to $p=0.77$ for the Gabor wavelet procedure. There is also an increase in the probability of no correlation for the maximum flux rates, from $p=0.17$ to $p=0.37$ for the GB02 procedure and $p=0.18$ to $p=0.91$ for the Gabor wavelet procedure.

\section{Discussion and Conclusions}

While the time of peak correlation between the mean phase travel times and the surface magnetic flux for most active regions occurs prior to the flux emergence, the correlation itself is not particularly strong. Some pixels may have a cross-correlation as strong as $0.35$, though the spatial average is generally much lower, on the order of $0.02$. Considering only pixels where the corresponding magnetic field is greater than $100$ G increases the spatial average by slightly more than a factor of two, to on the order of $0.05$. However, the non-uniformity of the correlation lag times about $T=0$ does seem to indicate that there is some degree of connection between the mean phase travel times and the emergence of active regions.

Examining the relationship between the correlation lag times and peak perturbation index times shows that, at least for the thresholds used in this work, the perturbation index itself is not a strong indicator of emerging magnetic flux. Of the 46 active regions studied, only 28 of these regions, on average, have the maximum perturbation index prior to the emergence of the magnetic flux. Additionally, only 30 of the active regions, on average, have the time of peak perturbation index preceding the correlation lag time. There is also no significant relationship between the time of peak perturbation index and maximum magnetic flux, maximum flux rate, or sunspot size, with the Spearman correlation coefficient $\rho$ on the order of $10^{-2}$.

In conclusion, we find that for a majority of active regions, the peak correlation between the surface magnetic field and the mean phase travel time deviations occurs before the active region emerges, with a median time lag in the best case of $-16$ hours prior to emergence. Separating the measurement annuli into additional arc segments does not necessarily increase the sensitivity of the measurements, and our results suggest that eight arc segments is optimal. While the median correlation lag time for the Gabor wavelet procedure is slightly greater than that of the GB02 procedure, the GB02 procedure has a greater number of active regions with a correlation lag time preceding the emergence of the active regions in every arc segment case. Additionally, we find no strong correlation between the lag times and sunspot size or maximum flux, though there is a noticeable correlation between the lag times and the maximum flux rate.

\bibliography{main}

\begin{thebibliography}{}
\expandafter\ifx\csname natexlab\endcsname\relax\def\natexlab#1{#1}\fi
\providecommand{\url}[1]{\href{#1}{#1}}
\providecommand{\dodoi}[1]{doi:~\href{http://doi.org/#1}{\nolinkurl{#1}}}
\providecommand{\doeprint}[1]{\href{http://ascl.net/#1}{\nolinkurl{http://ascl.net/#1}}}
\providecommand{\doarXiv}[1]{\href{https://arxiv.org/abs/#1}{\nolinkurl{https://arxiv.org/abs/#1}}}

\bibitem[{{Birch} {et~al.}(2013){Birch}, {Braun}, {Leka}, {Barnes}, \&
  {Javornik}}]{Birch_stats}
{Birch}, A.~C., {Braun}, D.~C., {Leka}, K.~D., {Barnes}, G., \& {Javornik}, B.
  2013, \apj, 762, 131, \dodoi{10.1088/0004-637X/762/2/131}

\bibitem[{Braun(2012)}]{Braun_comment}
Braun, D.~C. 2012, Science, 336, 296, \dodoi{10.1126/science.1215425}

\bibitem[{{Felipe} {et~al.}(2016){Felipe}, {Braun}, {Crouch}, \&
  {Birch}}]{Felipe}
{Felipe}, T., {Braun}, D.~C., {Crouch}, A.~D., \& {Birch}, A.~C. 2016, \apj,
  829, 67, \dodoi{10.3847/0004-637X/829/2/67}

\bibitem[{{Gizon} \& {Birch}(2002)}]{GB02}
{Gizon}, L., \& {Birch}, A.~C. 2002, \apj, 571, 966, \dodoi{10.1086/340015}

\bibitem[{{Hartlep} {et~al.}(2011){Hartlep}, {Kosovichev}, {Zhao}, \&
  {Mansour}}]{Hartlep}
{Hartlep}, T., {Kosovichev}, A.~G., {Zhao}, J., \& {Mansour}, N.~N. 2011,
  \solphys, 268, 321, \dodoi{10.1007/s11207-010-9544-1}

\bibitem[{{Ilonidis} {et~al.}(2013){Ilonidis}, {Zhao}, \&
  {Hartlep}}]{Stathis2013}
{Ilonidis}, S., {Zhao}, J., \& {Hartlep}, T. 2013, \apj, 777, 138,
  \dodoi{10.1088/0004-637X/777/2/138}

\bibitem[{{Ilonidis} {et~al.}(2011){Ilonidis}, {Zhao}, \&
  {Kosovichev}}]{Stathis_nature}
{Ilonidis}, S., {Zhao}, J., \& {Kosovichev}, A. 2011, Science, 333, 993,
  \dodoi{10.1126/science.1206253}

\bibitem[{Ilonidis {et~al.}(2012)Ilonidis, Zhao, \&
  Kosovichev}]{Ilonidis_response}
Ilonidis, S., Zhao, J., \& Kosovichev, A. 2012, Science, 336, 296,
  \dodoi{10.1126/science.1215539}

\bibitem[{{Kholikov}(2013)}]{Kholikov}
{Kholikov}, S. 2013, \solphys, 287, 229, \dodoi{10.1007/s11207-013-0321-9}

\bibitem[{{Korpi-Lagg} {et~al.}(2022){Korpi-Lagg}, {Korpi-Lagg}, {Olspert}, \&
  {Truong}}]{Korpi}
{Korpi-Lagg}, M.~J., {Korpi-Lagg}, A., {Olspert}, N., \& {Truong}, H.-L. 2022,
  arXiv e-prints, arXiv:2205.04419.
\newblock \doarXiv{2205.04419}

\bibitem[{{Moradi} {et~al.}(2009){Moradi}, {Hanasoge}, \& {Cally}}]{Moradi}
{Moradi}, H., {Hanasoge}, S.~M., \& {Cally}, P.~S. 2009, \apjl, 690, L72,
  \dodoi{10.1088/0004-637X/690/1/L72}

\bibitem[{{Nigam} {et~al.}(2007){Nigam}, {Kosovichev}, \&
  {Scherrer}}]{Gabor_nigam}
{Nigam}, R., {Kosovichev}, A.~G., \& {Scherrer}, P.~H. 2007, \apj, 659, 1736,
  \dodoi{10.1086/512535}

\bibitem[{{Singh} {et~al.}(2016){Singh}, {Raichur}, \& {Brandenburg}}]{Singh1}
{Singh}, N.~K., {Raichur}, H., \& {Brandenburg}, A. 2016, \apj, 832, 120,
  \dodoi{10.3847/0004-637X/832/2/120}

\bibitem[{{Stefan} {et~al.}(2021){Stefan}, {Kosovichev}, \&
  {Stejko}}]{StefanAR}
{Stefan}, J.~T., {Kosovichev}, A.~G., \& {Stejko}, A.~M. 2021, \apj, 913, 87,
  \dodoi{10.3847/1538-4357/abf2bf}

\bibitem[{{Toriumi} {et~al.}(2013){Toriumi}, {Ilonidis}, {Sekii}, \&
  {Yokoyama}}]{Toriumi}
{Toriumi}, S., {Ilonidis}, S., {Sekii}, T., \& {Yokoyama}, T. 2013, \apjl, 770,
  L11, \dodoi{10.1088/2041-8205/770/1/L11}

\bibitem[{{Waidele} {et~al.}(2022){Waidele}, {Roth}, {Singh}, \&
  {K{\"a}pyl{\"a}}}]{Waidele}
{Waidele}, M., {Roth}, M., {Singh}, N., \& {K{\"a}pyl{\"a}}, P. 2022, arXiv
  e-prints, arXiv:2202.11236.
\newblock \doarXiv{2202.11236}

\end{thebibliography}

\newpage

\newpage

\newpage

\newpage

\newpage

\end{document}